\documentclass[final]{svjour2}
\usepackage{graphicx,color}
\usepackage{rotating}
\usepackage{amssymb}
\usepackage{mathptmx}
\usepackage[numbers]{natbib}
\makeatletter
\journalname{Journal of Low Temperature Physics}

\bibpunct{}{}{,}{s}{}{,}

\begin{document}

\newcommand{\hdblarrow}{H\makebox[0.9ex][l]{$\downdownarrows$}-}


\title{Upper critical field measurements up to 60~T in arsenic-deficient 
LaO$_{0.9}$F$_{0.1}$FeAs$_{1-\delta}$: Pauli limiting behaviour at high fields 
vs.\ improved superconductivity at low fields}

\author{G.~Fuchs$^1$, S.-L.~Drechsler$^1$, N.~Kozlova$^1$, J.~Freudenberger$^1$,
M.~Bartkowiak$^2$, J.~Wosnitza$^2$, G.~Behr$^1$, K.~Nenkov$^1$, 
B.~B\"uchner$^1$, and L.~Schultz$^1$}

\institute{1:  Leibniz-Institut IFW Dresden, P.O.\ Box 270116, D-011171 
Dresden, Germany\\
Tel.: +49 351 4659 538\\ Fax: +49 351 4659 490\\
\email{fuchs@ifw-dresden.de}
\\2: Leibniz-Institut FZ Dresden-Rossendorf (FZD), Germany\\ 
}

\date{05.28.2009}

\maketitle

\begin{abstract}
We report resistivity and upper critical  field $B_{c2}(T)$
 data for As deficient LaO$_{0.9}$F$_{0.1}$FeAs$_{1-\delta}$  
in a wide temperature and high field range up to 60~T. These disordered 
samples exhibit a slightly enhanced  superconducting transition
at $T_c = 29$~K 
and a 
significantly enlarged slope d$B_{c2}$/d$T = -5.4$~T/K near $T_c$ 
which contrasts with a flattening of $B_{c2}(T)$ starting near 23~K above 
30~T. This flattening is interpreted as Pauli limiting behaviour (PLB) with 
$B_{c2}(0)\approx 63$~T. We compare our results with $B_{c2}(T)$-data 
reported in the literature for clean
 and disordered samples. Whereas clean samples show no PLB 
for fields below 60 to 70~T, the hitherto unexplained flattening of 
$B_{c2}(T)$
 for applied fields $H \parallel ab$ observed for several disordered closely 
related systems is interpreted also as a manifestation of PLB. 
Consequences of our results are discussed in terms of disorder effects within 
the frame of conventional and unconventional superconductivity. 
\end{abstract}



\vspace{0.4cm}
The recently discovered FeAs based superconductors \cite{Kamihara} exhibit high
transition temperatures $T_c \leq 57$~K and remarkably
 high upper critical fields $B_{c2}(0)$ exceeding often 70~T. Many basic 
properties 
of these novel superconductors and the underlying pairing mechanism are still 
not well understood. A study of $B_{c2}(T)$,
 in particular, 
investigations on disordered FeAs superconductors are of large interest since 
for an unconventional pairing both $T_c$ and d$B_{c2}$/d$T$ at $T_c$ are 
expected to be 
suppressed by  introducing 
disorder. In the present paper, 
$B_{c2}(T)$
of As-deficient LaO$_{0.9}$F$_{0.1}$FeAs$_{1-\delta}$ 
samples is studied in fields up to 60~T. 

Polycrystalline samples of LaO$_{0.9}$F$_{0.1}$FeAs were prepared by the 
standard solid state reaction method \cite{Zhu,Fuchs1}. Some samples have 
been wrapped in 
a Ta foil during the final annealing procedure. Ta acts as an As getter at 
high temperatures forming a solid solution of about 9.5 at.\% As in Ta. This 
leads to an As loss in the samples resulting in an As/Fe ratio of about 0.9. 
Due to disorder in the FeAs layer, an enhanced resistivity in the normal state 
at 31~K is found for the investigated As-deficient sample (ADS) exceeding that 
of a clean reference sample by a factor of about three. Nevertheless, the ADS 
has, with $T_c = 28.5$~K, a higher $T_c$ than stoichiometric reference 
samples ($T_c = 27.7$~K)\cite{Fuchs1}. 

	In Fig.\ 1, resistance data obtained in pulsed fields up to 50~T are 
shown for the ADS.  Gold contacts (100 nm thick) were prepared by sputtering 
in order to provide a low contact resistivity and, therefore, to avoid 
possible heating effects. The magnetic field generated by the employed IFW's 
pulsed field magnet rises within 10 ms to its maximum value $B_{max}$ 
and decreases 
afterwards to zero within the same time. The resistance data shown in Fig.\ 1 
were taken for descending field using $B_{max} = 47$~T. Additionally, 
resistance 
data were collected for $B_{max} = 29$~T at several selected temperatures. The 
agreement between the data obtained for both $B_{max}$-values confirms that 
our data are not affected by sample heating. 

	For polycrystalline samples, only the highest upper critical field 
$B_{c2}^{ab}$ is accessible which is related to 
those grains oriented with their 
$ab$-planes along the applied field. $B_{c2}^{ab}$ was determined 
from the onset 
of superconductivity (SC) defining 
\begin{figure}[bt]
\begin{center}
\includegraphics[width=0.72\linewidth, keepaspectratio]{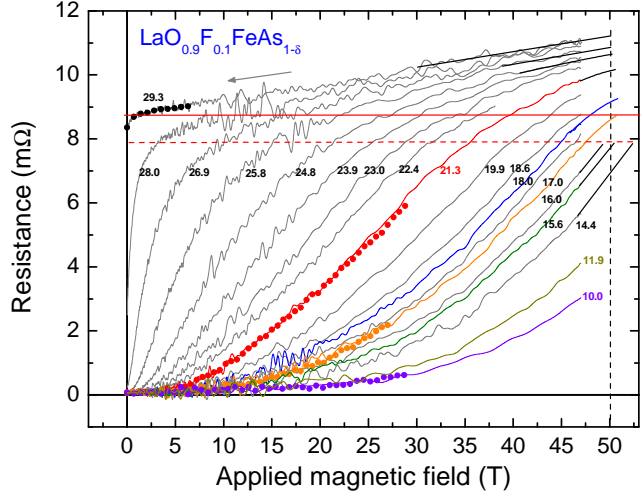}
\end{center}
\caption{(Color online)  Field dependence of the resistance at fixed $T$ 
(see legend) measured in pulsed fields. Lines: measurements up to 47~T; 
symbols measurements up to 29~T shown for selected $T$. Horizontal full 
and dashed lines: $R = R_N$ and $R = 0.9 R_N$, respectively with $R_N$ 
as the resistance in the normal state. }
\label{compare}
\end{figure}
it at 90\% of the resistance $R_N$ in the normal 
state (see Fig.\ 1). The temperature dependence of $B_{c2}^{ab}$ of our 
As-deficient sample obtained from pulsed field measurements in the IFW and the 
FZD is shown in Fig.\ 2 together with $B_{c2}$ data
reported for a clean reference sample \cite{Hunte}. The large slope 
d$B_{c2}/$d$T$= -5.4 T/K at $T_c$ of our ADS points to strong impurity 
scattering in accord with its enhanced resistivity at 30~K. For the clean 
sample \cite{Hunte} the available data up to 
45~T is well described by the WHH (Werthamer-Helfand-Hohenberg) 
model \cite{WHH}  
for the orbital limited upper critical field. Whereas for the ADS, the 
WHH model which predicts 
$B_{c2}^*(0) = 0.69T_c (\mbox{d}B_{c2}/\mbox{d}T)|_{T_c} = 106$~T at 
$T = 0$, fits the 
experimental data up to 30~T, only. For applied fields above 30~T increasing 
deviations from the WHH curve are clearly visible both for the $B_{c2}(T)$
 data from the IFW and the FZD. The flattening of $B_{c2}(T)$ at high field 
points to its limitation by the Pauli spin paramagnetism. This effect is 
measured in the WHH model by the Maki parameter 
$\alpha = \sqrt{2} B_{c2}^*(0)/B_p(0)$, where 
$B_p(0)$ is the Pauli limiting field. The paramagnetically limited upper 
critical field, $B_{c2}^p$, is  given by $B_{c2}^p(0) = B_{c2}^*(0)(1+\alpha^2)^{-0.5}$. For our ADS, a satisfying fit of the experimental data to this model was obtained for $\alpha = 1.31$ (see Fig.\ 2) 
and yields
$B_{c2}^p(0) = 63$~T. 
\begin{figure}
\begin{center}
\includegraphics[width=0.65\linewidth, keepaspectratio]{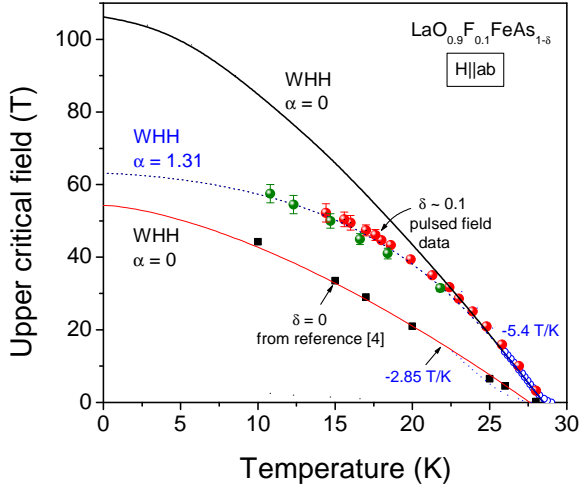}
\end{center}
\caption{(Color online) $T$-dependence of  $B_{c2}^{ab}$. Data for the 
As-deficient sample from 
DC ({\tiny $\blacksquare$}) and 
pulsed field measurements ( \textcolor{red}{$\bullet$}    - IFW Dresden,  
\textcolor{green}{$\bullet$}  - FZD) and data for a 
clean reference sample\cite{Hunte}. Solid lines: WHH model without PLB. 
Dotted line: $B_{c2}^*(T)$ for $\alpha = 1.31$ (see text).}
\label{compare}
\end{figure}

	For several disordered closely related systems, a similar flattening 
of $B_{c2}(T)$ as we found for our ADS has been reported for applied 
fields $H \parallel ab$. This is shown in Fig.\ 3 where the normalized 
upper critical field 
$h^* = B_{c2}(T)/\left[T_c(dB_{c2}/dT)|_{T_c}\right]$ is plotted against the reduced temperature 
$t = T/T_c$. In contrast, $B_{c2}(T)$ data for clean LaO$_{0.93}$F$_{0.07}$FeAs 
samples \cite{Kohama} (see Fig.\ 3) show almost no Pauli-limiting behavior for fields up to 70 T. The 
data in Fig.\ 3
are well described by the WHH model using the obtained Maki parameters 
$\alpha$. 
The deviation of $h^*(t)$ at low $T$ from $h^*(t)$  for $\alpha = 0$ increases with $\alpha$
 due to rising paramagnetic pair-breaking.

	We found for our As-deficient samples indications for a 
strongly enhanced Pauli paramagnetism from $\mu$SR experiments 
\cite{Fuchs2}. Their improved SC at 
high $T$
and low fields can be understood within conventional s-wave SC by enhanced 
disorder. In contrast, for clean FeAs superconductors an unconventional 
$s^{\pm}$-wave 
scenario has been proposed. On the basis of our results for $B_{c2}(T)$, two 
alternative scenarios of opposite disorder effects might be suggested: 
(i) an impurity-driven change of the pairing state from $s^{\pm}$ to 
conventional $s_{++}$-wave SC and (ii) a 
special impurity-driven stabilization of the $s^{\pm}$ state where the 
As-vacancies are assumed to scatter predominantly within the bands. 
The PLB found here suggests to continue measurements at least 
up to 70~T in order to eludicate, whether there is still much room for 
increasing 
$B_{c2}$ beyond that range. The possibility to improve the low-field 
properties of FeAs superconductors by introducting As vacancies 
opens new preparation routes for optimising the properties of these 
superconductors.
\begin{figure}
\begin{center}
\includegraphics[width=0.65\linewidth, keepaspectratio]{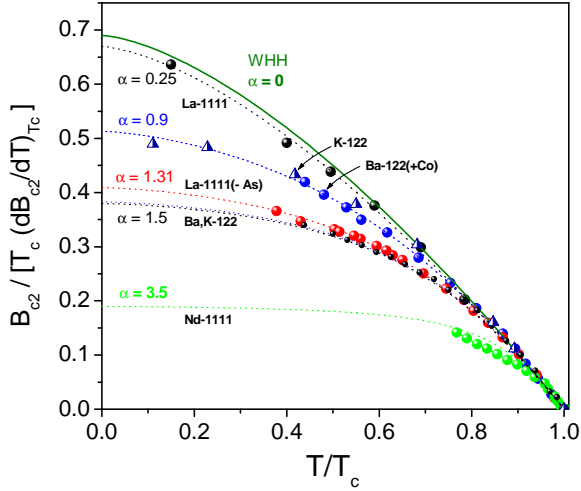}
\end{center}
\caption{(Color online) Normalized upper critical field 
$B_{c2}(T)/\left[T_c\left(\mbox{d}B_{c2}/\mbox{d}T\right)|_{T_c}\right]$ 
vs.\ $T/T_c$ 
for an As-deficient LaO$_{0.9}$F$_{0.1}$FeAs$_{1-\delta}$ sample 
(La-1111(-As)) in comparison with data reported for non-deficient 
LaO$_{0.93}$F$_{0.07}$FeAs (La-1111, $T_c = 25$~K) \cite{Kohama}, 
Ba(Fe$_{0.9}$Co$_{0.1}$)$_2$As$_2$ (Ba-122(+Co), $T_c = 21.9$~K) 
\cite{Yamamoto}, KFe$_2$As$_2$ (K-122, $T_c = 2.8$~K)\cite{Terashima}, 
Ba$_{0.55}$K$_{0.45}$Fe$_2$As$_2$ 
(Ba-122, $T_c = 32$~K)\cite{Altarawneh}, 
NdO$_{0.7}$F$_{0.3}$FeAs 
(Nd-1111, $T_c = 45.6$~K)\cite{Jaroszynski}. Dotted 
and solid lines: WHH model for the 
indicated  values. All curves shown correspond to $H \parallel ab$.  }
\label{compare}
\end{figure}


\end{document}